\documentclass[letter]{aa} 
\usepackage{graphicx}
\usepackage[varg]{txfonts}
\usepackage[citecolor=blue, colorlinks=true]{hyperref}
\bibstyle{aa}
\newcommand{\magethanks}{Based on observations obtained at the Magellan I (Baade)
Telescope at Las Campanas Observatory, Chile.}

\def\arc{PSZ1-ARC\,G311.6602$-$18.4624}
\def\lya{Ly$\alpha$}
\def\kms{km s$^{-1}$}
\def\LyA{Lyman $\alpha$}

\begin{document} 

\title{The Sunburst Arc: Direct Lyman $\alpha$ escape observed in the
	brightest known lensed galaxy\thanks{\magethanks}}

\titlerunning{Direct Ly$\alpha$ escape from the brightest lensed arc in the Universe}

\author{
	T. E. Rivera-Thorsen\inst{1}
\and
	H. Dahle\inst{1}
\and
	M. Gronke\inst{1}
\and
	M. Bayliss\inst{2}
\and
	J. R. Rigby\inst{3}
\and
	R. Simcoe\inst{2}
\and
	R. Bordoloi\inst{4,2}
\and
        M. Turner\inst{2,5}
\and
        G. Furesz\inst{2}
} 

\institute{
        Institute of Theoretical Astrophysics, 
	University of Oslo, Postboks 1029, 0315 Oslo, Norway\\
	\email{eriveth@astro.uio.no}
	\and
        MIT-Kavli Center for Astrophysics and Space Research, 
	77 Massachusetts Avenue, Cambridge, MA, 02139, USA
	\and
	Observational Cosmology Lab, NASA Goddard Space Flight Center, 
	8800 Greenbelt Rd., Greenbelt, MD 20771, USA
	\and
	Hubble Fellow
	\and
	Las Cumbres Observatory, 6740 Cortona Dr, Goleta, CA 93117, USA
}

\date{Received: 26 October 2017, Accepted: 20 November 2017}

\abstract{

	We present rest-frame ultraviolet and optical spectroscopy of the
	brightest lensed galaxy yet discovered, at redshift $z=2.4$. The 
	source reveals a characteristic, triple-peaked Lyman $\alpha$ profile
	which has been predicted in various theoretical works but to our
	knowledge has not been unambiguously observed previously. The feature is
	well fit by a superposition of two components: a double-peak profile
	emerging from substantial radiative transfer, and a narrow, central
	component resulting from directly escaping \LyA{} photons; but is poorly
	fit by either component alone. We demonstrate that the feature is
	unlikely to contain contamination from nearby sources, and that the
	central peak is unaffected by radiative transfer effects apart from very
	slight absorption. The feature is detected at signal-to-noise ratios
	exceeding 80 per pixel at line center, and bears strong resemblance to
	synthetic profiles predicted by numerical models.  

}

\authorrunning{T.~E.~Rivera-Thorsen et al.}


\maketitle

\section{Introduction}

Young, star-forming galaxies are generally believed to be the most important
source of the Lyman continuum (LyC) radiation responsible for the reionization
of the early Universe \citep[e.g.][]{Bouwens2012,Faisst2016}. However, faint
galaxies of high star formation rates tend to contain large amounts of neutral
Hydrogen, which is opaque to LyC at column densities $\log(N)\gtrsim 17.2$
\citep[e.g.][]{Osterbrock}. Exactly where in the galaxy this radiation
originates, and how it finds its way into the intergalactic medium, remains one
of the most important unsolved problems in Astrophysics. 

Lyman continuum can be allowed to escape a galaxy mainly by two means.
In one scenario, the source regions can be surrounded by a largely isotropic gas
with column density of neutral Hydrogen $N_{\mathrm{HI}}$ sufficiently low to
not completely attenuate the passing LyC radiation. This is usually
referred to as the \emph{density bounded medium}. In a second scenario, the
neutral medium is optically thick, but not completely covering the ionizing
sources. This scenario has been referred to as the \emph{picket fence model}
\citep[e.g.][]{Concelice2000,Heckman2011}, the \emph{riddled ionization-bounded
medium} \citep{Verhamme2015}, or the \emph{ionization-bounded medium with holes}
\citep{Zackrisson2013}. The latter scenario can be subdivided into two
cases: One quasi-isotropic, where the medium consists of clumps with a
combined covering fraction below unity. This scenario will potentially have a
large number of narrow, direct lines of sight (LOS) in random directions
\citep{HansenOh2006,Duval2014,Gronke2016Letter}. In the other, the medium is
closer to an optically thick shell, perforated in few places by channels of very
low optical depth \citep[e.g.][]{Zackrisson2013,Behrens2014}. The clumpy and the
perforated shell cases are not always well distinguished in the
literature; we here reserve the term \emph{picket fence model} for the clumpy
scenario.  While these three scenarios have similar observational signatures in
LyC, they have different spectral signatures in \LyA{}, which  is shaped by
resonant interactions with the \ion{H}{i} that also governs escape of LyC,
although in different ways \citep[see e.g.][and references
therein]{DijkstraRev}. 

\begin{figure}[t] 
	\centering
	\includegraphics[width=\columnwidth]{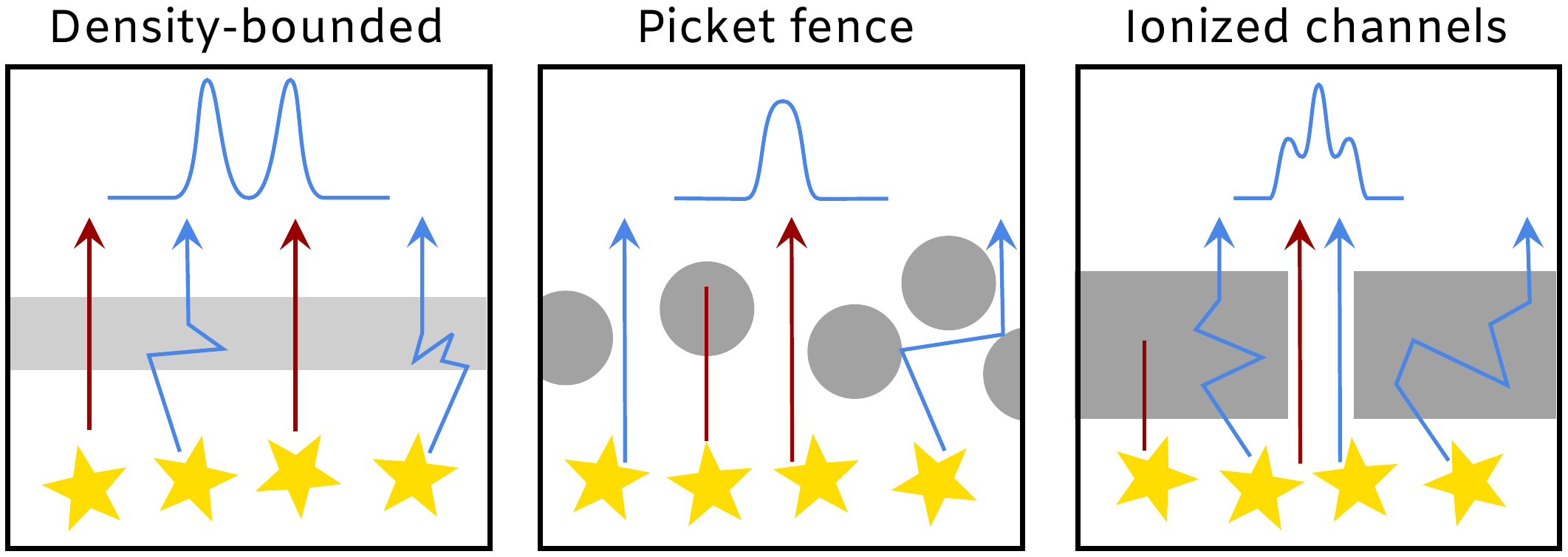}
	\caption{Cartoons showing a fully
		covering $\lesssim 10^{17}\,\mathrm{cm}^{-2}$ \ion{H}{i} screen 
		(\textbf{left}), the ``picket fence''
		scenario of a clumpy medium with many ionized sight lines and
		weak RT effects in \lya{} (\textbf{center}), and the scenario with
		few ionized channels through a neutral medium optically thick
		in both LyC and \lya{} (\textbf{right}). Blue arrows show
	\lya{}, red arrows LyC.\label{fig:gascartoon}}
\end{figure}

Figure~\ref{fig:gascartoon} shows a cartoon depiction of these three scenarios
and their emerging \lya{} spectral shapes. Blue arrows signify
\lya{}, and red arrows LyC radiation. The \emph{left panel} shows the density
bounded model.  \LyA{}, which has an interaction cross section $\sim 3$ orders
of magnitude higher at line center, undergoes substantial radiative transfer
before escaping as a broadened, double-peaked profile, while LyC escapes due to
the low optical depth of the medium \citep{Jaskot2013,JaskotOey,Verhamme2015}.
Most local LyC leakers display \lya{} line shapes similar to this
\citep{Verhamme2015}, and also in metal absorption are more consistent with this
scenario \citep{Chisholm2017}. The \emph{central panel} shows the picket fence
model, in which LyC can escape only through open sight lines between dense
clumps, while \lya{} can bounce off boundaries of clumps and thus escape after
few scatterings and only weak radiative transfer effects. The resulting \lya{}
spectrum is a narrow, single peaked line, with the intrinsic shape nearly intact
\citep[e.g.][]{Gronke2016Letter}. The \emph{right panel} shows the perforated
shell model. Here, LyC and a central \lya{} peak are also escaping directly
through an open channel. In addition to this, \lya{} photons are captured
scattering in the denser neutral gas, leading to substantial radiative transfer
effects. The resulting \lya{} line shape is a narrow, central peak escaping
through the uncovered sight lines, superimposed on the characteristic double
peak profile emerging from the optically thick
gas\citep[e.g.][]{Duval2014,Behrens2014,Gronke2016Letter}. \cite{Herenz2017}
report a possible observation of such ionized channels seen from the side in the
nearby galaxy SBS 0335$-$052E.

In this letter, we present rest-frame UV and optical spectroscopy of the lensed
galaxy \arc{}, which we nickname the Sunburst Arc\footnote{A sunburst being a
``direct view at the sun through rifted clouds''.}. We find that it contains
strong evidence for direct escape of Lyman $\alpha$ photons through a
perforated, optically thick medium. The galaxy was serendipitously discovered in
follow-up imaging of its lensing cluster, which was discovered through its
Sunyaev-Zel'dovich effect in the Planck data \citep{PlanckSZ2014} and was first
reported and described by~\cite{Dahle2016}. This is to our knowledge the first
unambiguous detection of this type of spectral signature, and it bears
remarkable resemblance to theoretical predictions of e.g.~\cite{Behrens2014}. 

\section{Observations and data reduction}

\begin{figure}[t]
	\includegraphics[width=.95\columnwidth]{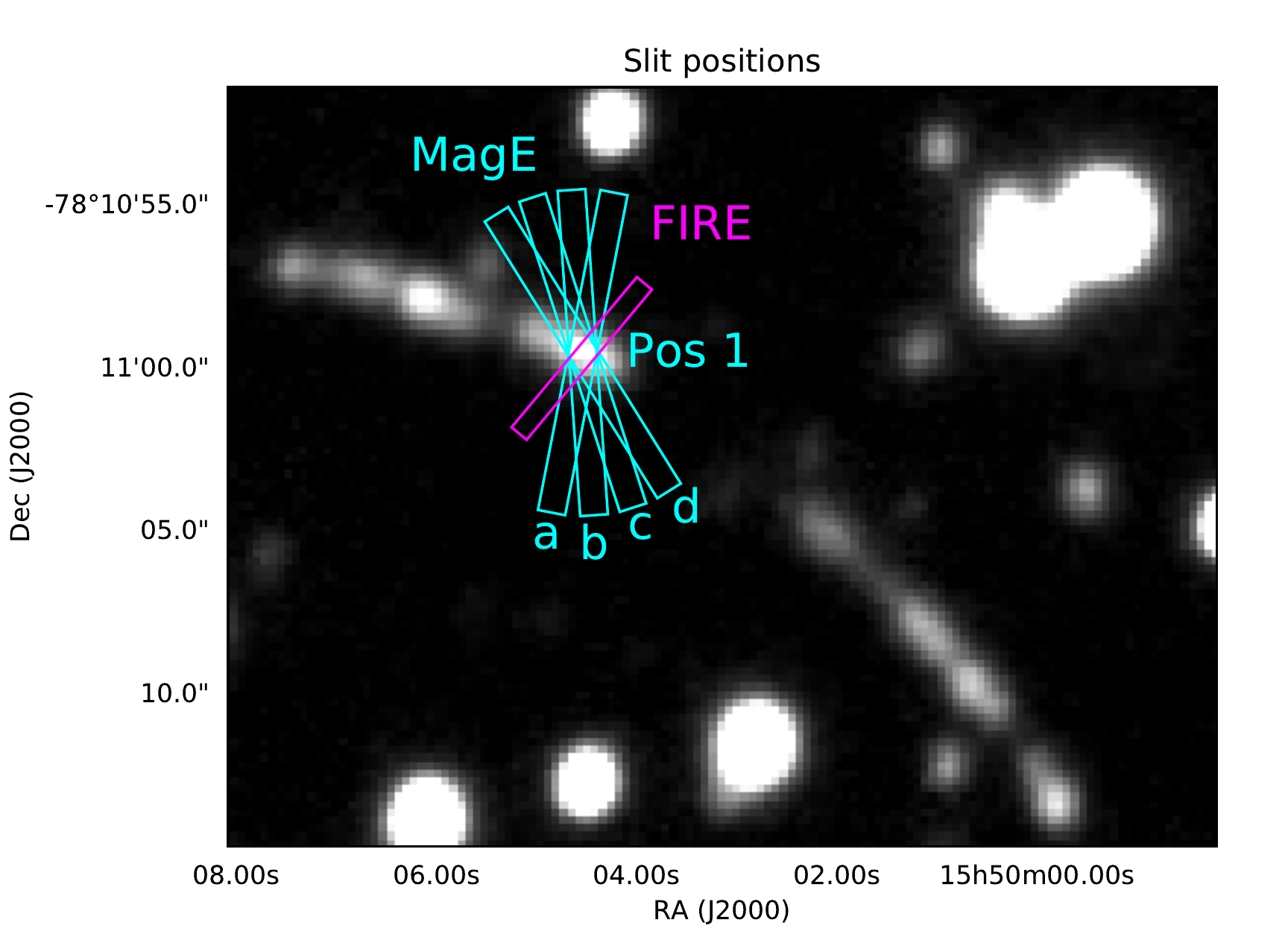} \caption{NTT/EFOSC2
R-band image of \arc, with MagE (cyan) and FIRE (magenta) slit pointings
overlaid. N is up, E is left.\label{fig:finderchart}} \end{figure}

We observed \arc{} through variable cloud cover on UT 24 May 2017, beginning at
03:31, with the Magellan Echellette (MagE) spectrograph \citep{MagE2008} on the
Magellan-I (Baade) telescope. MagE was configured with the 1\farcs0 slit, which
delivers a constant spectral resolution as a function of wavelength,
R$\simeq4700$ ($\sim60$ km s$^{-1}$) as measured by Gaussian fits to night sky
lines. We performed an alternating sequence of four 2700s science frames and 6s
reference arc lamp calibration frames, with the slit position angle being
rotated between exposures to match the parallactic angle at the midpoint of each
science frame (see Fig.~\ref{fig:finderchart}). The seeing varied between
$\sim$1\farcs1--1\farcs5, and the cloud cover increased significantly over the
course of the observations (see Fig.~\ref{fig:lya2d}). 
We also observed a flux calibrator star, CD-23d \citep{Hamuy1992,Hamuy1994}, at
three different airmasses bracketing those of the science observations, as well
as a second standard, EG131, for correcting telluric absorption lines. The MagE
spectra were processed using the MagE pipeline, which is part of the Carnegie
Python Distribution\footnote{Described at http://code.obs.carnegiescience.edu/},
and were fluxed as described in \cite{RigbyMagasauraI}.  Due to the 
variable cloud cover we only recover a relative flux calibration as a function of
wavelength, with no constraint on the absolute flux normalization.

We also observed the arc on UT 30 March 2016, beginning at 09:06 hrs, with the
Folded-port InfraRed Echelle (FIRE) spectrograph \citep{Fire2013} on the Baade
telescope. Observations were performed with the echelle grating and the 0\farcs6
slit as a pair of 1800s exposures with the target nodded between two positions
along the slit, and the slit aligned perpendicular to the arc (see
Fig.~\ref{fig:finderchart}). Conditions were clear throughout the observations,
with seeing averaging $\sim$0\farcs8. We also observed standard star HIP77712
immediately before the science observations to provide a flux calibration and
telluric corrections. The data were reduced using the FireHose v2 data reduction
pipeline \citep{Firehose2015}. The final reduced spectrum has a spectral
resolution of R$\simeq6000$ ($\sim50$ km s$^{-1}$).  The wavelength
ranges covered are $3200 - 8200$ Å for MagE and $0.82 - 2.49 \mu$ for FIRE.\@ A 
more comprehensive analysis of these and additional observations is to be 
presented in a subsequent paper (Rivera-Thorsen et al., in prep).
 
We determined the systemic redshift of the arc based on the FIRE spectrum by
fitting a single Gaussian profile to each of the four strong emission lines
H$\beta$, [\ion{O}{III}]$\lambda\lambda$ (4959, 5007), and H$\alpha$. We then
computed an uncertainty-weighted average of the four redshifts to be
$z_{\mathrm{neb}}=2.37094 \pm 0.00001$.


\section{Results}

\subsection{Triple-peaked \LyA{} emission}

\begin{figure}[t]
	\includegraphics[width=\columnwidth]{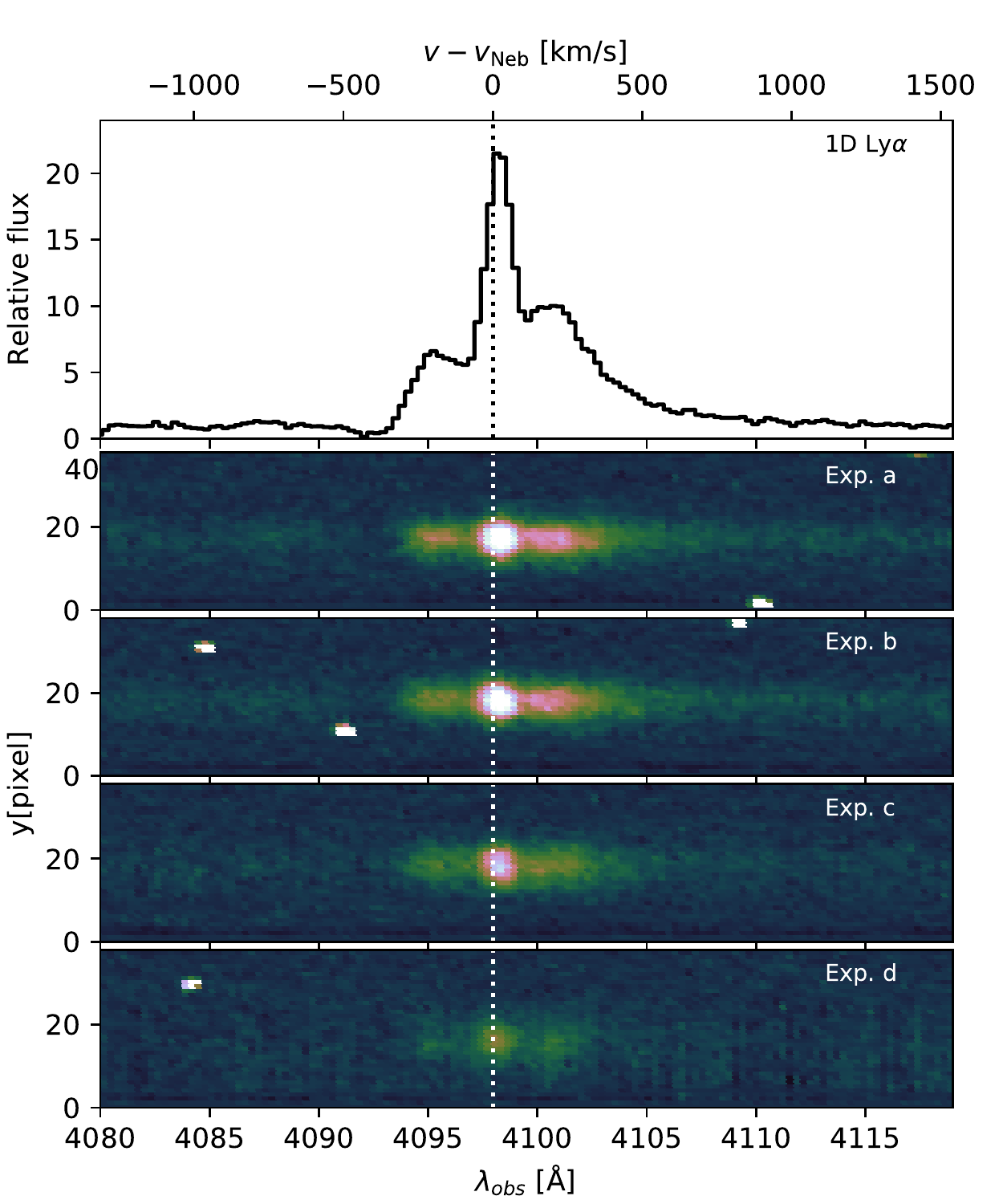}
	\caption{\textbf{Top:} Observed \lya{} in black. The error spectrum, not
		shown, is
		comparable to the line width. \textbf{Lower panels:}
		Two-dimensional Ly$\alpha$ profiles of the 4 MagE
		exposures. Color scale is linear, and cut levels are set to
		enhance detail. The vertical dotted line indicates the best-fit
                velocity zero point from rest frame optical nebular 
	transitions.\label{fig:lya2d}} 
\end{figure}

Figure~\ref{fig:lya2d} shows cutouts of \lya{} from the reduced 2D MagE spectra
of the four pointings shown in Fig.~\ref{fig:finderchart}. The top panel
contains the coadded and extracted 1D profile on the same scale.
The bright, narrow peak at $v\approx0$ km s$^{-1}$ is immediately evident for
all of the position angles, and the strongly similar line morphologies point to
the flux originating in the same spatial region, at least down to the scale of
the seeing. The 1D profile is a textbook example of the kind of line shape we
would expect to observe when looking through a clear channel in an otherwise
optically thick neutral medium; consisting of a double-peaked profile typical
for the \lya{} profile emerging through a large number of scatterings with a
narrow, non-scattered emission peak at line center superimposed onto it, as
shown in e.g. Fig.~7 of~\cite{Behrens2014}.  Note that \cite{Behrens2014} have
not systematically evaluated their models in a region of parameter
space, so a quantitative comparison is not possible at the moment, but we plan
to model this in Rivera-Thorsen et al., in prep.  It is however clear
that unlike the typical optically thin, density-bounded scenario, this profile
also shows strong signs of radiative transfer in at least a modestly thick,
surrounding neutral medium.

\subsection{Lyman $\alpha$ radiative transfer modeling}

\begin{figure} 
	\centering
	\includegraphics[width=\columnwidth]{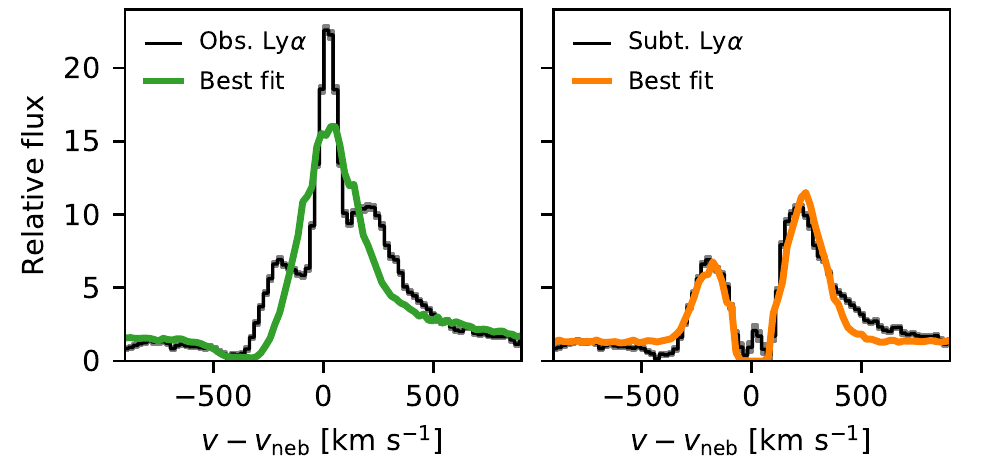}
        \caption{Best fits of expanding, isotropic shell models to the observed
		data (\textbf{left}) and with the central peak subtracted
		(\textbf{right}).\label{fig:shellmodel}} 
\end{figure}

To test whether the observed profile is consistent with the perforated neutral
shell scenario, we have attempted to fit the grid of spherically symmetric,
isotropic expanding shell models presented in~\cite{Gronke2015} \citep[inspired
by similar models by ][]{Ahn2003,Verhamme2006,Schaerer2011}, to the observed
\lya{} profile (see Fig.~\ref{fig:shellmodel}). As a prior to the intrinsic
profile, we used the width and position of observed H$\alpha$. We first
attempted to fit the full, observed \lya{} profile to the shell model grid
(\emph{left panel}), but could not even approximately reproduce it, even though
shell models are able to reproduce the vast majority of \lya{} emission profiles
remarkably well \citep[e.g.][]{Gronke2017MUSE}. We then fit the central, narrow
peak to a single Gaussian profile, which was subtracted, and the remaining
profile, similar to typical, double-peaked \lya{} profiles, was again fitted to
the shell model grid (\emph{right panel}).  Unlike the full observed
profile, we were able to reproduce the approximated double-peak profile. The
fit should not be taken too literally, as the subtraction of the central peak is
a coarse approximation, but it does support the hypothesis of a double-peak
profile resulting from substantial radiative transfer, with a largely unaltered,
intrinsic  component superimposed upon it.

We then compare the narrow central peak to the shape of H$\alpha$ to test
whether this component is consistent with direct escape.  If no neutral hydrogen
is encountered along the LOS, these two profiles will be identical up to a
multiplicative constant. Significant amounts of \ion{H}{I}, on the other hand,
will result in an altered \lya{} line shape, broadened by frequency diffusion.
The upper panel in Fig.~\ref{fig:lineslineup} shows for comparison the
normalized \lya{} and H$\alpha$ profiles. H$\alpha$, observed with FIRE, has
been smoothed to match the poorer instrument resolution of MagE and scaled to
match the height of \lya{}.  The profiles of H$\alpha$ and the central \lya{}
peak follow each other very well on the red side, as would be expected from the
perforated, optically thick scenario.  Surprisingly, \lya{} is narrower than the
intrinsic line shape on the blue side, suggesting a weak interaction with
neutral gas, absorbing a minor fraction of the light. We interpret this as being
due to a residual neutral component in diffuse ionized interstellar or
circumgalactic gas along the LOS and note that the low-velocity \ion{Si}{iv}
feature in Fig.~\ref{fig:lineslineup} lines up well with the blue edge of the
central \lya{} peak. 

\subsection{Interstellar metal lines}

\begin{figure}[t]
	\includegraphics[width=\columnwidth]{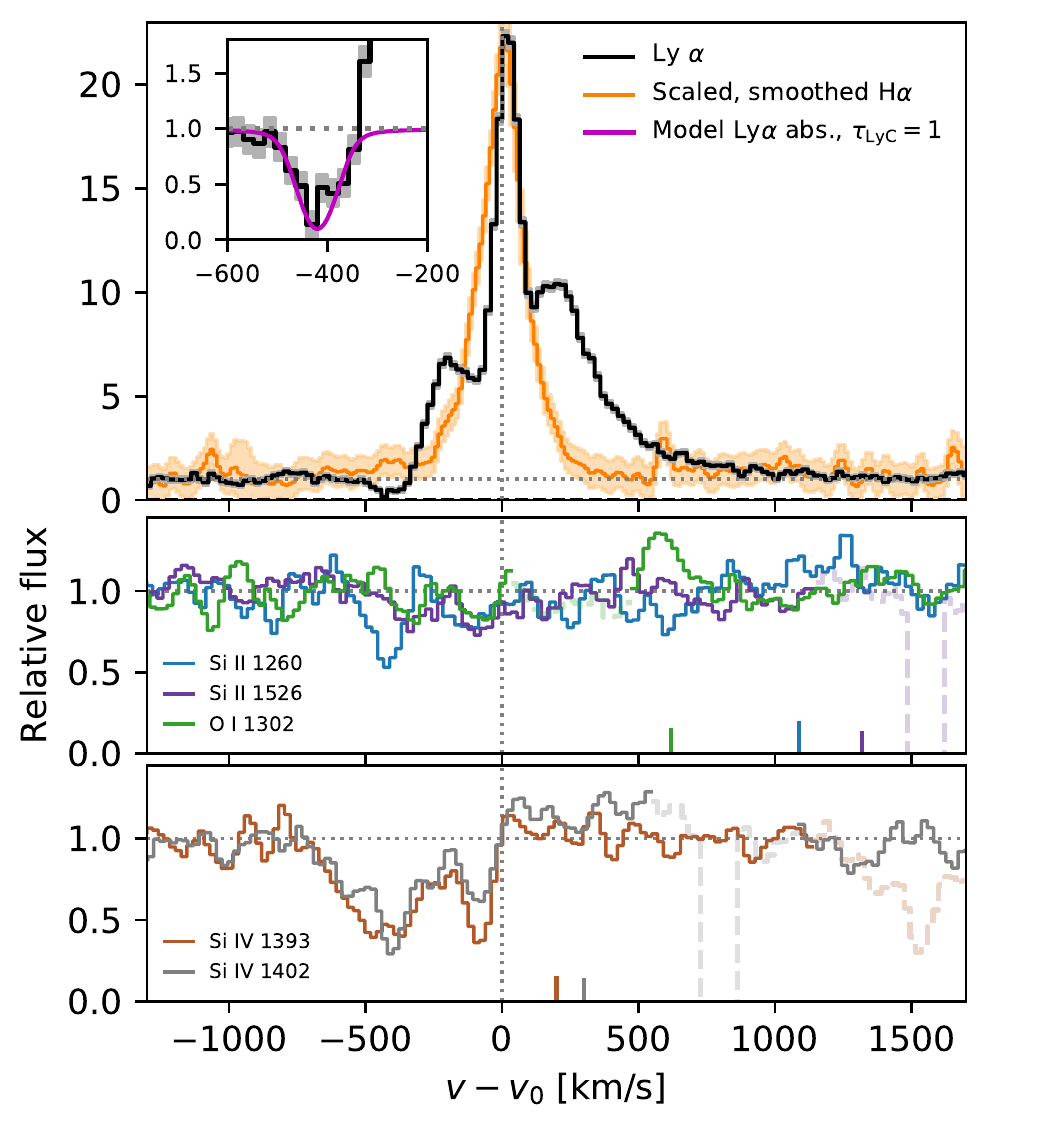}
	\caption{Continuum normalized line profiles. \textbf{Upper panel:}
		\LyA{} and smoothed, scaled H$\alpha$ (see text for details).
		\emph{Inset:} Zoom-in on \lya{}, with the theoretical absorption
		profile for $\tau_{\mathrm{LyC}}=1$ superimposed.
		\textbf{Middle panel:} Selected low-ionization metal absorption.
		Colored markers show the approximate expected position of
		fluorescent emission lines, color coded by their corresponding
		resonant line, with lengths indicating the average error
		spectrum in a velocity range of $\pm 1500$ km~s$^{-1}$ around
		the line.\textbf{Lower panel:}  \ion{Si}{IV}. Colored markers
		show the average errors, but are placed at arbitrary
		wavelengths.  Metal lines are smoothed by 3 px. 
		Gray dotted lines denote continuum level (\emph{horizontal})
		and velocity zero point (\emph{vertical}).  Pale dashed lines show
                regions masked out due to contamination.\label{fig:lineslineup}} 
\end{figure}

The lower two panels of Fig.~\ref{fig:lineslineup} show a selection of
metal lines of the neutral and ionized phase. These lines were
normalized by fitting a linear function to the local continuum and dividing by
the resulting line. We have omitted transitions blueward of rest frame \lya{} due
to \lya{} forest absorption. The shown data has been smoothed by a boxcar kernel
of 3 pixels width. Two main components of absorbing gas are evident in the
\ion{Si}{IV} spectra; one at low velocity ($v \sim -100$ km s$^{-1}$), and one at
intermediate velocity ($v \sim -420$ km s$^{-1}$).  The neutral line profiles are
much shallower but follow the general morphology of the ionized medium,
reflecting that the majority of ionized gas is likely to be found in the same
cloud systems as the neutral gas, on the side facing the central ionizing
source.

The neutral lines in the low-velocity component have approximately the
same depth, suggesting that the gas is concentrated in optically thick clumps of
low velocity-binned covering fraction \citep[see e.g.][]{RiveraThorsen2017}. 
In contrast, the strong and weak absorption lines behave very differently in the
$v \sim -420$ \kms{} component, with the weak lines being either non-detections
or extremely shallow.  A quantitative analysis of this is to be made in our
follow-up paper, but this shows that at least in the blue end, the
neutral gas is optically thin, at least to the weaker transitions, and thus
might cover the background fully. At this velocity, it has little
impact on the line shape of \lya{}, but may still affect LyC escape
substantially.

All the ions shown in the middle panel of Fig.~\ref{fig:lineslineup} have fine
structure split ground levels and thus some amount of fluorescent emission is
expected. As argued by~\cite{JaskotOey}, a comparison of resonant absorption
depth and fluorescent emission strength can hint at the possible presence of
neutral gas within the spectrograph aperture, but off the LOS.\@ We have shown
the approximate expected centroids of fluorescent emission in
Fig.~\ref{fig:lineslineup}. Despite very shallow absorption, at least some of
these lines show significant fluorescent emission, suggesting the presence of
significant amounts of off-LOS neutral gas, in agreement with what is expected
from the radiative transfer effects seen in \lya{}.

\subsection{Conditions for Lyman continuum escape}

The presence of empty or highly ionized lines of sight makes the Sunburst Arc a
prime candidate for spatially resolved observation of LyC. Some uncertainty
arises from the component at $v\sim -420$ km s$^{-1}$. Looking at
Fig.~\ref{fig:lineslineup}, the \ion{Si}{ii} lines have strongly differing
depth, implying that this component is optically thin at least in $\lambda 1526$
and possibly also in $\lambda 1260$, in which case the neutral gas component could be
covering the entire background source. Due to the high velocity, this has little
impact on \lya{} escape, but LyC is sensitive to neutral gas at any velocity and
could possibly be blocked by this system. A full analysis of this is planned for
a follow-up paper, but as a first test, we have modelled the theoretical \lya{}
absorption profile arising from a gas component at $v=-420$ \kms{} with
$N_{\ion{H}{i}}=10^{17.2}$ cm$^{-2}$ (corresponding to $\tau_{\mathrm{LyC}}=1$,
$f_{\mathrm{esc}}(\mathrm{LyC}) \sim 0.4$))
and $T=10^4$K. The resulting profile, convolved with the MagE instrument
profile, is shown in magenta in the inset of Fig.~\ref{fig:lineslineup}. The
theoretical line is too strong to be consistent with the observed feature. We 
tentatively conclude that the neutral component at $v \sim -420$ km s$^{-1}$
has either too low column density or covering fraction to effectively block
LyC escape. 

An HST mid-cycle proposal to image the arc in rest-frame LyC has recently been
accepted.

\section{Summary}

We have presented Magellan/MagE and Magellan/FIRE spectra of the extremely
bright, strongly lensed galaxy \arc{}, which we nickname the ``Sunburst Arc''.
By comparing the 2D spectra of four different position angles, we find it is
unlikely that any contamination from nearby objects is present.  Based on strong
Balmer- and [\ion{O}{III}] emission lines, we have improved the precision of
previous redshift estimates. From this, we have found that the Ly$\alpha$ line
shape in the combined MagE spectrum is a close match to theoretical predictions
for direct Ly$\alpha$ escape through a perforated neutral medium. By comparing
to the H$\alpha$ emission feature, we have found that most likely, this narrow,
central peak has undergone no other radiative transfer than very slight
absorption, probably by a residual neutral fraction of \ion{H}{I} in the ionized
medium along the open line of sight. The observed line profile cannot be
modeled by isotropic expanding shell models, but a reasonable fit can be
obtained by removing the central peak. A medium-velocity neutral gas component
at $v\sim-420$ km s$^{-1}$ causes some uncertainty as to whether the
LOS seen in Ly$\alpha$ would also be transparent in LyC, but the
absorption feature it leaves in the Ly$\alpha$ continuum suggests it is either
optically thin to LyC or not fully covering.

\begin{acknowledgements} We thank the anonymous referee for thorough and
	constructive suggestions, and E. C. Herenz for useful comments and
	ideas.  ER-T thanks Stockholm University for their kind hospitality.
	ER-T and HD acknowledge support from the Research Council of Norway.
\end{acknowledgements}

\bibliographystyle{aa}
\bibliography{main}

\begin{thebibliography}{30}
\expandafter\ifx\csname natexlab\endcsname\relax\def\natexlab#1{#1}\fi

\bibitem[{{Ahn} {et~al.}(2003){Ahn}, {Lee}, \& {Lee}}]{Ahn2003}
{Ahn}, S.-H., {Lee}, H.-W., \& {Lee}, H.~M. 2003, \mnras, 340, 863

\bibitem[{{Behrens} {et~al.}(2014){Behrens}, {Dijkstra}, \&
  {Niemeyer}}]{Behrens2014}
{Behrens}, C., {Dijkstra}, M., \& {Niemeyer}, J.~C. 2014, \aap, 563, A77

\bibitem[{{Bouwens} {et~al.}(2012){Bouwens}, {Illingworth}, {Oesch}, {Trenti},
  {Labb{\'e}}, {Franx}, {Stiavelli}, {Carollo}, {van Dokkum}, \&
  {Magee}}]{Bouwens2012}
{Bouwens}, R.~J., {Illingworth}, G.~D., {Oesch}, P.~A., {et~al.} 2012, \apjl,
  752, L5

\bibitem[{{Chisholm} {et~al.}(2017){Chisholm}, {Orlitov{\'a}}, {Schaerer},
  {Verhamme}, {Worseck}, {Izotov}, {Thuan}, \& {Guseva}}]{Chisholm2017}
{Chisholm}, J., {Orlitov{\'a}}, I., {Schaerer}, D., {et~al.} 2017, \aap, 605,
  A67

\bibitem[{{Conselice} {et~al.}(2000){Conselice}, {Gallagher}, {Calzetti},
  {Homeier}, \& {Kinney}}]{Concelice2000}
{Conselice}, C.~J., {Gallagher}, J.~S., {Calzetti}, D., {Homeier}, N., \&
  {Kinney}, A. 2000, \aj, 119, 79

\bibitem[{{Dahle} {et~al.}(2016){Dahle}, {Aghanim}, {Guennou}, {Hudelot},
  {Kneissl}, {Pointecouteau}, {Beelen}, {Bayliss}, {Douspis}, {Nesvadba},
  {Hempel}, {Gronke}, {Burenin}, {Dole}, {Harrison}, {Mazzotta}, \&
  {Sunyaev}}]{Dahle2016}
{Dahle}, H., {Aghanim}, N., {Guennou}, L., {et~al.} 2016, \aap, 590, L4

\bibitem[{{Dijkstra}(2014)}]{DijkstraRev}
{Dijkstra}, M. 2014, \pasa, 31, 40

\bibitem[{{Duval} {et~al.}(2014){Duval}, {Schaerer}, {{\"O}stlin}, \&
  {Laursen}}]{Duval2014}
{Duval}, F., {Schaerer}, D., {{\"O}stlin}, G., \& {Laursen}, P. 2014, \aap,
  562, A52

\bibitem[{{Faisst}(2016)}]{Faisst2016}
{Faisst}, A.~L. 2016, \apj, 829, 99

\bibitem[{{Gagné} {et~al.}(2015){Gagné}, Lambrides, Faherty, \&
  Simcoe}]{Firehose2015}
{Gagné}, J., Lambrides, E., Faherty, J.~K., \& Simcoe, R. 2015, Firehose v2.0

\bibitem[{{Gronke}(2017)}]{Gronke2017MUSE}
{Gronke}, M. 2017, ArXiv e-prints [\eprint[arXiv]{1709.07008}]

\bibitem[{{Gronke} {et~al.}(2015){Gronke}, {Bull}, \& {Dijkstra}}]{Gronke2015}
{Gronke}, M., {Bull}, P., \& {Dijkstra}, M. 2015, \apj, 812, 123

\bibitem[{{Gronke} {et~al.}(2016){Gronke}, {Dijkstra}, {McCourt}, \&
  {Oh}}]{Gronke2016Letter}
{Gronke}, M., {Dijkstra}, M., {McCourt}, M., \& {Oh}, S.~P. 2016, \apjl, 833,
  L26

\bibitem[{{Hamuy} {et~al.}(1994){Hamuy}, {Suntzeff}, {Heathcote}, {Walker},
  {Gigoux}, \& {Phillips}}]{Hamuy1994}
{Hamuy}, M., {Suntzeff}, N.~B., {Heathcote}, S.~R., {et~al.} 1994, \pasp, 106,
  566

\bibitem[{{Hamuy} {et~al.}(1992){Hamuy}, {Walker}, {Suntzeff}, {Gigoux},
  {Heathcote}, \& {Phillips}}]{Hamuy1992}
{Hamuy}, M., {Walker}, A.~R., {Suntzeff}, N.~B., {et~al.} 1992, \pasp, 104, 533

\bibitem[{{Hansen} \& {Oh}(2006)}]{HansenOh2006}
{Hansen}, M. \& {Oh}, S.~P. 2006, \mnras, 367, 979

\bibitem[{{Heckman} {et~al.}(2011){Heckman}, {Borthakur}, {Overzier},
  {Kauffmann}, {Basu-Zych}, {Leitherer}, {Sembach}, {Martin}, {Rich},
  {Schiminovich}, \& {Seibert}}]{Heckman2011}
{Heckman}, T.~M., {Borthakur}, S., {Overzier}, R., {et~al.} 2011, \apj, 730, 5

\bibitem[{{Herenz} {et~al.}(2017){Herenz}, {Hayes}, {Papaderos}, {Cannon},
  {Bik}, {Melinder}, \& {{\"O}stlin}}]{Herenz2017}
{Herenz}, E.~C., {Hayes}, M., {Papaderos}, P., {et~al.} 2017, ArXiv e-prints
  [\eprint[arXiv]{1708.07007}]

\bibitem[{{Jaskot} \& {Oey}(2013)}]{Jaskot2013}
{Jaskot}, A.~E. \& {Oey}, M.~S. 2013, \apj, 766, 91

\bibitem[{{Jaskot} \& {Oey}(2014)}]{JaskotOey}
{Jaskot}, A.~E. \& {Oey}, M.~S. 2014, ArXiv e-prints
  [\eprint[arXiv]{1406.4413}]

\bibitem[{{Marshall} {et~al.}(2008){Marshall}, {Burles}, {Thompson},
  {Shectman}, {Bigelow}, {Burley}, {Birk}, {Estrada}, {Jones}, {Smith},
  {Kowal}, {Castillo}, {Storts}, \& {Ortiz}}]{MagE2008}
{Marshall}, J.~L., {Burles}, S., {Thompson}, I.~B., {et~al.} 2008, in
  \procspie, Vol. 7014, Ground-based and Airborne Instrumentation for Astronomy
  II, 701454

\bibitem[{{Osterbrock} \& {Ferland}(2006)}]{Osterbrock}
{Osterbrock}, D.~E. \& {Ferland}, G.~J. 2006, {Astrophysics of gaseous nebulae
  and active galactic nuclei}

\bibitem[{{Planck Collaboration} {et~al.}(2014){Planck Collaboration}, {Ade},
  {Aghanim}, {Armitage-Caplan}, {Arnaud}, {Ashdown}, {Atrio-Barandela},
  {Aumont}, {Aussel}, {Baccigalupi}, \& et~al.}]{PlanckSZ2014}
{Planck Collaboration}, {Ade}, P.~A.~R., {Aghanim}, N., {et~al.} 2014, \aap,
  571, A29

\bibitem[{{Rigby} {et~al.}(2017){Rigby}, {Bayliss}, {Sharon}, {Gladders},
  {Chisholm}, {Dahle}, {Johnson}, {Paterno-Mahler}, {Wuyts}, \&
  {Kelson}}]{RigbyMagasauraI}
{Rigby}, J.~R., {Bayliss}, M.~B., {Sharon}, K., {et~al.} 2017, ArXiv e-prints
  [\eprint[arXiv]{1710.07294}]

\bibitem[{{Rivera-Thorsen} {et~al.}(2017){Rivera-Thorsen}, {{\"O}stlin},
  {Hayes}, \& {Puschnig}}]{RiveraThorsen2017}
{Rivera-Thorsen}, T.~E., {{\"O}stlin}, G., {Hayes}, M., \& {Puschnig}, J. 2017,
  \apj, 837, 29

\bibitem[{{Schaerer} {et~al.}(2011){Schaerer}, {Hayes}, {Verhamme}, \&
  {Teyssier}}]{Schaerer2011}
{Schaerer}, D., {Hayes}, M., {Verhamme}, A., \& {Teyssier}, R. 2011, \aap, 531,
  A12

\bibitem[{{Simcoe} {et~al.}(2013){Simcoe}, {Burgasser}, {Schechter}, {Fishner},
  {Bernstein}, {Bigelow}, {Pipher}, {Forrest}, {McMurtry}, {Smith}, \&
  {Bochanski}}]{Fire2013}
{Simcoe}, R.~A., {Burgasser}, A.~J., {Schechter}, P.~L., {et~al.} 2013, \pasp,
  125, 270

\bibitem[{{Verhamme} {et~al.}(2015){Verhamme}, {Orlitov{\'a}}, {Schaerer}, \&
  {Hayes}}]{Verhamme2015}
{Verhamme}, A., {Orlitov{\'a}}, I., {Schaerer}, D., \& {Hayes}, M. 2015, \aap,
  578, A7

\bibitem[{{Verhamme} {et~al.}(2006){Verhamme}, {Schaerer}, \&
  {Maselli}}]{Verhamme2006}
{Verhamme}, A., {Schaerer}, D., \& {Maselli}, A. 2006, \aap, 460, 397

\bibitem[{{Zackrisson} {et~al.}(2013){Zackrisson}, {Inoue}, \&
  {Jensen}}]{Zackrisson2013}
{Zackrisson}, E., {Inoue}, A.~K., \& {Jensen}, H. 2013, \apj, 777, 39

\end{thebibliography}
\end{document}